\documentstyle[psfig]{aipproc}

\begin{document}

\title{Monte Carlo methods and applications for the nuclear shell model}

\author{D.J. Dean$^*$ and J.A. White$^\dagger$}

\address{$^*$Physics Division, Oak Ridge National Laboratory,
Oak Ridge, TN \\
and Physics Department, University of Tennessee, Knoxville, TN \\
$^\dagger$W. K. Kellogg Radiation Laboratory, 106-38, California
Institute of Technology \\ Pasadena, California 91125 USA}


\maketitle

\begin{abstract}
The shell-model Monte Carlo (SMMC) technique transforms the
traditional nuclear shell-model problem into a path-integral
over auxiliary fields. We describe below the method and its
applications to four physics issues: calculations of
$sd$-$pf$- shell nuclei, a discussion of
electron-capture rates in $pf$-shell nuclei,
exploration of pairing correlations in unstable nuclei, and
level densities in rare earth systems.
\end{abstract}

\section{Introduction}

Studies of nuclei far from stability have long been a goal
of nuclear science. Nuclei on either side of the stability
region, either neutron-rich or deficient,
are being produced at new radioactive beam facilities across the
world. At these facilities, and with the help of
advances in nuclear many-body theory, the community
will address many of the key physics issues including:
mapping of the neutron and proton drip lines, thus exploring the limits
of stability;
understanding effects of the
continuum on weakly bound nuclear systems;
understanding the nature of shell gap modifications in
very neutron-rich systems;
determining nuclear properties needed for astrophysics;
investigating deformation, spin, and pairing properties of systems
far from stability; and analyzing microscopically
unusual shapes in unstable nuclei.

The range and diversity of nuclear behavior, as indicated in the
above list of ongoing and planned experimental investigations, have
naturally engendered a host of theoretical models.
Short of a complete solution to the
many-nucleon problem, the interacting shell model is
widely regarded as the most broadly capable description of
low-energy nuclear structure, and the one most directly
traceable to the fundamental many-body problem.
Difficult though it may be, solving the shell-model problem
is of fundamental importance to our understanding of the
correlations found in nuclei.

One avenue of research during the past few years has been in the area of
the nuclear shell model solved not by diagonalization, but by integration.
In what follows, we will describe the shell-model Monte Carlo (SMMC)
method and discuss several
recent and interesting results obtained from theory. These
include calculations in $sd$-$pf$-shell neutron-rich nuclei,
a discussion of electron-capture rates in
$fp$-shell nuclei, pairing correlations in medium-mass nuclei near N$=$Z,
and studies of level densities in rare-earth nuclei.

\section{SMMC Methods}

In the following we briefly outline the formalism of the SMMC method. We
begin with a brief description of statistical mechanics techniques used in
our approach, then
discuss the Hubbard-Stratonovich transformation, and end with a
discussion of Monte Carlo sampling procedures. We refer the reader to
previous works \cite{Lang,Koonin97} for a more detailed exposition.

\subsection{Observables}

SMMC methods rely on an ability to calculate the imaginary-time many-body
evolution operator, $\exp (-\beta H)$, where $\beta$ is a real
$c$-number. The many-body Hamiltonian can be written schematically as
\begin{equation}
 H=\varepsilon {\cal O}
+{1\over2}V {\cal O} {\cal O}\;,
\label{eq_a}
\end{equation}
where ${\cal O}$ is a density operator, $V$ is
the strength of the two-body interaction, and $\varepsilon$ is a
single-particle energy. In the full problem, there are many such
quantities with various orbital indices that are summed over, but we omit
them here for the sake of clarity.

While the SMMC technique does not result in a complete solution to the
many-body problem in the sense of giving all eigenvalues and eigenstates of
$ H$, it can result in much useful information. For example, the
expectation value of some observable, $\Omega$, can be obtained by
calculating
\begin{equation}
\langle  \Omega\rangle= {{\rm Tr}\,
e^{-\beta H} \Omega\over {\rm Tr}\,
e^{-\beta H}}\;.
\label{eq_b}
\end{equation}
Here, $\beta\equiv T^{-1}$ is interpreted as the inverse of the temperature
$T$, and the many-body trace is defined as
\begin{equation}
{\rm Tr}\, X\equiv\sum_i \langle i\vert  X\vert i\rangle\;,
\label{eq_c}
\end{equation}
where the sum is over many-body states of the system. In the canonical
ensemble, this sum is over all states with a specified number of nucleons
(implemented by number projection \cite{Ormand94,Koonin97}), while the grand
canonical ensemble introduces a chemical potential and sums over {\it all}
many-body states.

In the limit of low temperature ($T\rightarrow0$ or
$\beta\rightarrow\infty$), the canonical trace reduces to a ground-state
expectation value. Alternatively, if $\vert \Phi\rangle$ is a many-body trial
state not orthogonal to the exact ground state, $\vert\Psi\rangle$, then
$e^{-\beta H}$ can be used as a filter to refine $\vert\Phi\rangle$ to
$\vert\Psi\rangle$ as $\beta$ becomes large. An observable can be calculated
in this ``zero temperature'' method as
\begin{equation}
{\langle\Phi\vert e^{-{\beta\over2} H} \Omega
e^{-{\beta\over2} H} \vert\Phi\rangle\over
\langle\Phi\vert e^{-\beta H}\vert\Phi\rangle}
{}~\hbox{$\longrightarrow\hskip-20pt{}^{\beta\rightarrow\infty}$}
{\langle\Psi\vert \Omega\vert\Psi\rangle\over
\langle\Psi\vert\Psi\rangle}\;.
\label{eq_d}
\end{equation}
If $ \Omega$ is the Hamiltonian, then (\ref{eq_d}) at $\beta=0$ is the
variational estimate of the energy, and improves as $\beta$ increases. Of
course, the efficiency of the refinement for any observable depends upon the
degree to which $\vert\Phi\rangle$ approximates $\vert\Psi\rangle$.

Beyond such static properties, $e^{-\beta H}$ allows us to obtain some
information about the dynamical response of the system. For an operator
$\Omega$, the response function, $R_\Omega(\tau)$, in the canonical
ensemble is defined as
\begin{equation}
R_\Omega(\tau)\equiv {{\rm Tr}\,
e^{-(\beta-\tau) H}  \Omega^\dagger
e^{-\tau H} \Omega\over {\rm Tr}\,
e^{-\beta H}}\equiv \langle\Omega^\dagger(\tau)\Omega(0)\rangle,
\label{eq_e}
\end{equation}
where $\Omega^\dagger(\tau)\equiv e^{\tau H} \Omega^\dagger
e^{-\tau H}$ is the imaginary-time Heisenberg operator. Interesting
choices for $\Omega$ are the annihiliation operators
for particular orbitals, the
Gamow-Teller, $M1$, or quadrupole moment, etc. Inserting complete sets of
$A$-body eigenstates of $ H$ ($\{\vert i\rangle,\vert f\rangle\}$ with
energies $E_{i,f}$) shows that
\begin{equation}
R_\Omega(\tau) ={1\over Z}\sum_{if}
e^{-\beta E_i} \vert\langle f\vert\Omega\vert i\rangle\vert^2
e^{-\tau(E_f-E_i)},
\label{eq_f}
\end{equation}
where $Z=\sum_i e^{-\beta E_i}$ is the partition function. Thus,
$R_\Omega(\tau)$ is the Laplace transform of the strength function
$S_\Omega(E)$:
\begin{eqnarray}
R_\Omega(\tau)& = & \int^\infty_{-\infty} e^{-\tau E} S_\Omega(E)dE\;; \\
S_\Omega(E)&=& {1\over Z}\sum_{fi} e^{-\beta E_i} \vert\langle f\vert
\Omega\vert i\rangle\vert^2 \delta(E-E_f+E_i)\;.
\label{eq_g}
\end{eqnarray}

\noindent
Hence, if we can calculate $R_\Omega(\tau)$, $S_\Omega(E)$ can be determined.
Short of a full inversion of the Laplace transform (which is often
numerically difficult), the behavior of $R_\Omega(\tau)$ for small $\tau$
gives information about the energy-weighted moments of $S_\Omega$. In
particular,
\begin{equation}
R_\Omega(0)=\int^\infty_{-\infty} S_\Omega (E) dE=
{1\over Z}\sum_i e^{-\beta E_i} \vert\langle f\vert {\Omega}\vert
i\rangle\vert^2=
\langle {\Omega}^\dagger{\Omega}\rangle_A
\label{eq_h}
\end{equation}
is the total strength,
\begin{equation} -R_\Omega^\prime (0)=
\int^\infty_{-\infty} S_\Omega (E) EdE=
{1\over Z}\sum_{if} e^{-\beta E_i} \vert\langle f\vert \Omega\vert
i\rangle\vert^2 (E_f-E_i)
\label{eq_i}
\end{equation}
is the first moment (the prime denotes differentiation with respect to
$\tau$).

It is important to note that we usually cannot obtain detailed spectroscopic
information from SMMC calculations. Rather, we can calculate expectation
values of operators in the thermodynamic ensembles or the ground state.
Occasionally, these can indirectly furnish properties of excited states. For
example, if there is a collective $2^+$ state absorbing most of the $E2$
strength, then the centroid of the quadrupole response function will be a
good estimate of its energy. But, in general, we are without the numerous
specific excitation energies and wave functions that characterize a direct
diagonalization. This is both a blessing and a curse. The former is that for
the very large model spaces of interest, there is no way in which we can deal
explicitly with all of the wave functions and excitation energies. Indeed, we
often don't need to, as experiments only measure average nuclear properties
at a given excitation energy. The curse is that comparison with detailed
properties of specific levels is difficult. In this sense, the SMMC method is
complementary to direct diagonalization for modest model spaces, but is the
only method for treating very large problems.

\subsection{The Hubbard-Stratonovich transformation}

It remains to describe the Hubbard-Stratonovich ``trick'' by which
$e^{-\beta H}$ is managed. In broad terms, the difficult many-body
evolution is replaced by a superposition of an infinity of tractable one-body
evolutions, each in a different external field, $\sigma$. Integration over
the external fields then reduces the many-body problem to quadrature.

To illustrate the approach, let us assume that only one operator ${\cal
O}$ appears in the Hamiltonian (\ref{eq_a}).  Then all of the difficulty
arises from the two-body interaction, that term in $ H$ quadratic in
${\cal O}$. If $ H$ were solely linear in ${\cal O}$, we would
have a one-body quantum system, which is readily dealt with. To linearize the
evolution, we employ the Gaussian identity
\begin{equation}
e^{-\beta H}=
\sqrt{\beta \mid V\mid \over 2\pi} \int^\infty_{-\infty} d\sigma
e^{-{1\over2}\beta \mid V\mid \sigma^2}
e^{-\beta h};\;\;\;
 h= \varepsilon {\cal O} +s V\sigma{\cal O}\;.
\label{eq_j}
\end{equation}
Here, $ h$ is a one-body operator associated with a $c$-number field
$\sigma$, and the many-body evolution is obtained by integrating the one-body
evolution, $ U_\sigma\equiv e^{-\beta h}$, over all $\sigma$ with a
Gaussian weight. The phase, $s$, is $1$ if $V<0$, or $i$ if $V>0$.
Equation~(\ref{eq_j}) is easily verified by completing the square in the
exponent of the integrand; since we have assumed
that there is only a single operator ${\cal O}$, there is no need to
worry about non-commutation.

For a realistic Hamiltonian, there will be many non-commuting density
operators, ${\cal O}_\alpha$, present, but we can always reduce
the two-body term to diagonal form. Thus for a general two-body interaction
in a general time-reversal invariant form, we write
\begin{equation}
{H}=\sum_\alpha \left(\epsilon^\ast_\alpha {\bar{\cal O}}_\alpha+
\epsilon_\alpha  {\cal O}_\alpha\right)+
{1\over2}\sum_\alpha V_\alpha \left\{  {\cal O}_\alpha,
 {\bar{\cal O}}_\alpha\right\}\;,
\label{eq_k}
\end{equation}
where $ {\bar{\cal O}}_\alpha$ is the time reverse of $ {\cal
O}_\alpha$. Since, in general, $[ {\cal O}_\alpha, {\cal
O}_\beta]\not=0$, we must split the interval $\beta$ into $N_t$ ``time
slices'' of length $\Delta\beta\equiv\beta/N_t$,
\begin{equation}
e^{-\beta H}= [e^{-\Delta\beta H}]^{N_t}, \label{eq_l}
\end{equation}
and for each time slice $n=1, \ldots, N_t$ perform a linearization similar to
Eq.~\ref{eq_j} using auxiliary fields $\sigma_{\alpha n}$. Note that because
the various $ {\cal O}_\alpha$ need not commute, the representation of
$e^{-\Delta\beta h}$ must be accurate through order $(\Delta\beta)^2$ to
achieve an overall accuracy of order $\Delta\beta$.

We are now able to write expressions for observables as the ratio of two
field integrals. Thus expectations of observables can be written as
\begin{eqnarray}
\langle \Omega\rangle =
{\int{\cal D} \sigma W_\sigma \Omega_\sigma\over
\int{\cal D} \sigma W_\sigma},\qquad\qquad\qquad\qquad \label{eq_m} \\
\noalign{\noindent where\hfill}
W_\sigma = G_\sigma {\rm Tr}\, U_\sigma\;;\qquad G_\sigma=
e^{-{\Delta\beta}\sum_{\alpha n}\mid V_\alpha\mid \mid\sigma_{\alpha
n}\mid^2}\;;\nonumber \\
\Omega_\sigma= {{\rm Tr}\,  U_\sigma  \Omega\over {\rm Tr}\,
U_\sigma}\;;\qquad
{\cal D} \sigma \equiv \prod^{N_t}_{n=1}\prod_\alpha d\sigma_{\alpha
n}d\sigma_{\alpha n}^{*}
\left(\Delta\beta\vert V_\alpha\vert\over 2\pi\right), \label{eq_n} \\
\noalign{\noindent and\hfill}
 U_\sigma=  U_{N_t}\ldots  U_2 U_1\;;\qquad
 U_n= e^{-\Delta\beta h_n};\nonumber \\
 h_n =\sum_\alpha \left(\varepsilon_\alpha^{*} +
s_\alpha V_\alpha \sigma_{\alpha n}\right){\bar{\cal O}}_\alpha+
\left(\varepsilon_\alpha +
s_\alpha V_\alpha \sigma_{\alpha n}^{*}\right){\cal O}_\alpha\;.
\label{eq_o}
\end{eqnarray}

\noindent
This is, of course, a discrete version of a path integral over $\sigma$.
Because there is a field variable for each operator at each time slice, the
dimension of the integrals ${\cal D} \sigma$ can be very large, often
exceeding $10^5$. The errors in Eq.~\ref{eq_m} are of order $\Delta\beta$, so
that high accuracy requires large $N_t$ and perhaps extrapolation to
$N_t=\infty$ ($\Delta\beta=0$).

Thus, the many-body observable is the weighted average (weight $W_\sigma$) of
the observable $\Omega_\sigma$ calculated in an ensemble involving only the
one-body evolution $ U_\sigma$. Similar expressions involving two
$\sigma$ fields (one each for $e^{-\tau H}$ and $e^{-(\beta-\tau)
H}$) can be written down for the response function (\ref{eq_e}), and all are
readily adapted to the canonical or grand canonical ensembles or to the
zero-temperature case.

An expression of the form (\ref{eq_m}) has a number of attractive features.
First, the problem has been reduced to quadrature---we need only calculate
the ratio of two integrals. Second, all of the quantum mechanics (which
appears in $\Omega_\sigma$) is of the one-body variety, which is simply
handled by the algebra of $N_s\times N_s$ matrices. The price to pay is that
we must treat the one-body problem for all possible $\sigma$ fields.

\subsection{Monte Carlo quadrature and the sign problem}

We employ the Metropolis, Rosenbluth, Rosenbluth, Teller, and Teller
algorithm \cite{Metropolis} to generate the field configurations,
$\sigma$, which requires only the ability to calculate the weight
function for a given value of the integration variables.  This method
requires that the weight function $W_\sigma$ must be real and
non-negative. Unfortunately, many of the Hamiltonians of physical
interest suffer from a sign problem, in that
$W_\sigma$ is negative over significant fractions of the integration volume.
The fractional variance of a given expectation value becomes unacceptably
large as the average sign approaches zero.

It was shown \cite{Koonin97}
that for even-even and $N=Z$ nuclei there is no
sign problem for Hamiltonians if all $V_\alpha\leq 0$.  Such forces include
reasonable approximations to the realistic Hamiltonian like
pairing-plus-multipole interactions.  However, for an arbitrary
Hamiltonian, we are not guaranteed that all $V_\alpha \leq 0$ (see, for
example,  Alhassid {\it et al.} \cite{Alhassid}).
However, we may expect that a {\it realistic} Hamiltonian will be dominated
by terms like those of the schematic pairing-plus-multipole force (which is,
after all, why the schematic forces were developed) so that it is, in some
sense, close to a Hamiltonian for which the MC is directly applicable. Thus,
the ``practical solution'' to the sign problem presented in
Alhassid {\em et al.}~\cite{Alhassid}
is based on an extrapolation of observables calculated
for a ``nearby'' family of Hamiltonians whose integrands have a positive
sign. Success depends crucially upon the degree of extrapolation required.
Empirically, one finds that for all of the many realistic interactions
tested in the $sd$- and $pf$-shells, the extrapolation required is modest,
amounting to a factor-of-two variation in the isovector monopole pairing
strength.

Based on the above observation, it is possible to decompose $ H$ in
Eq.~\ref{eq_k} into its ``good'' and ``bad'' parts, ${ H}= { H}_G+
{ H}_B$.
The ``good'' Hamiltonian, ${ H}_G$, includes, in addition to the one-body
terms, all the two-body interactions with $V_\alpha \leq0$, while the ``bad''
Hamiltonian, ${ H}_B$, contains all interactions with $V_\alpha>0$. By
construction, calculations with ${ H}_G$ alone have $\Phi_\sigma\equiv1$
and are thus free of the sign problem.

We define a family of Hamiltonians, ${ H}_g$, that depend on a continuous
real parameter~$g$ as ${ H}_g=f(g){ H}_G+g { H}_B$, so that
${ H}_{g=1}={ H}$, and $f(g)$ is a function with $f(1)=1$ and
$f(g<0)>0$ that can be chosen to make the extrapolations less severe. (In
practical applications, $f(g)=1-(1-g)/\chi$ with $\chi\approx4$, and
applied only to the two-body terms in $H_G$ has been found
to be a good choice.) If the $V_\alpha$ that are large in magnitude are
``good,'' we expect that ${ H}_{g=0}={ H}_G$ is a reasonable starting
point for the calculation of an observable $\langle{{ \Omega} }\rangle$.
One might then hope to calculate $\langle{{ \Omega}}\rangle_g={\rm
Tr}\,({ \Omega}e^{-\beta  H_g})/{\rm Tr}\,(e^{-\beta  H_g})$ for
small $g>0$ and then to extrapolate to $g=1$, but typically
$\langle\Phi\rangle$ collapses even for small positive $g$. However, it is
evident from our construction that ${ H}_g$ is characterized by
$\Phi_\sigma\equiv1$ for any $g\leq 0$, since all the ``bad'' $V_\alpha(>0)$
are replaced by ``good'' $g V_\alpha<0$. We can therefore calculate
$\langle{{ \Omega}}\rangle_g$ for any $g\leq0$ by a Monte Carlo sampling
that is free of the sign problem. If $\langle{{ \Omega} }\rangle_g$ is a
smooth function of $g$, it should then be possible to extrapolate to $g=1$
(i.e., to the original Hamiltonian) from $g\leq0$. We emphasize that $g=0$ is
not expected to be a singular point of $\langle{{ \Omega} }\rangle_g$; it
is special only in the Monte Carlo evaluation. The extrapolation methods
we employ have been tested against standard shell-model diagonalizations
in many cases, and
have, in general, been shown to work very well \cite{Koonin97}.

\section{Applications}

\subsection{$sd$-$pf$ nuclei}

Studies of extremely neutron-rich nuclei have revealed a number of
intriguing new phenomena.  Two sets of these nuclei that have received
particular attention are those with neutron number $N$ in the vicinity
of the $1s0d$ and $0f_{7/2}$ shell closures ($N \approx 20$ and $N
\approx 28$).  Experimental studies of neutron-rich Mg and Na isotopes
indicate the onset of deformation, as well as the modification of the
$N = 20$ shell gap for $^{32}$Mg and nearby nuclei \cite{r:motobayashi}.
Inspired by the rich set of phenomena occurring near the $N = 20$
shell closure when $N \gg Z$, attention has been directed to nuclei
near the $N = 28$ (sub)shell closure for a number of S and Ar isotopes
\cite{r:brown1,r:brown2} where similar, but less dramatic, effects
have been seen as well.

In parallel with the experimental efforts, there have been several
theoretical studies seeking to understand and, in some cases, predict
properties of these unstable nuclei.  Both mean-field
\cite{r:werner,r:campi} and shell-model calculations
\cite{r:brown1,r:brown2,r:wbmb,r:poves1,r:fukunishi,r:retamosa,r:caurier}
have been proposed. The latter require a severe truncation to
achieve tractable model spaces, since the successful description of these
nuclei involves active nucleons in both the $sd$- and the $pf$-shells.
The natural basis for the problem is therefore the full $sd$-$pf$
space, which puts it out of reach of exact diagonalization on current
hardware.

SMMC methods offer an alternative to direct
diagonalization when the bases become very large. Though SMMC provides
limited detailed spectroscopic information, it can predict, with good
accuracy, overall nuclear properties such as masses, total strengths,
strength distributions, and deformation --- precisely those quantities
probed by the recent experiments. It thus seems natural to apply SMMC
methods to these unstable neutron-rich nuclei. Two questions will
arise --- center-of-mass motion and choice of the interaction --- that
are not exactly new, but demand special treatment in very large spaces.
These questions were addressed in detail in Ref.~\cite{dean_cm}. We present
a brief selection of results here.

There is limited experimental information about the highly unstable,
neutron-rich nuclei under consideration.  In many cases only the mass,
excitation energy of the first excited state, the $B(E2)$ to that state,
and the $\beta$-decay rate is known, and not even all of this
information is available in some cases.  From the
measured $B(E2)$, an estimate of the nuclear deformation parameter,
$\beta_2$, has been obtained via the usual relation
\begin{equation}
\beta_2 = 4 \pi \sqrt{B(E2; 0^+_{gs} \rightarrow 2^+_1)}/3 Z R_0^2 e
\end{equation}
with $R_0 = 1.2 A^{1/3}$ fm and $B(E2)$ given in $e^2$fm$^4$.

Much of the interest in the region stems from the unexpectedly large
values of the deduced $\beta_2$, results which suggest the onset of
deformation and have led to speculations about the vanishing of the $N
= 20$ and $N = 28$ shell gaps.  The lowering in energy of the 2$^+_1$
state supports this interpretation.  The most thoroughly studied case,
and the one which most convincingly demonstrates these phenomena, is
$^{32}$Mg with its extremely large $B(E2) = 454 \pm 78 \, e^2$fm$^4$ and
corresponding $\beta_2 = 0.513$ \cite{r:motobayashi}; however, a word of
caution is necessary when deciding on the basis of this
limited information that we are in the presence of well-deformed
rotors: for $^{22}$Mg, we would obtain $\beta_2 = 0.67$, even more
spectacular, and for $^{12}$C, $\beta_2 = 0.8$, well above the
superdeformed bands.

\begin{figure}
\hskip 1.5in {\psfig{file=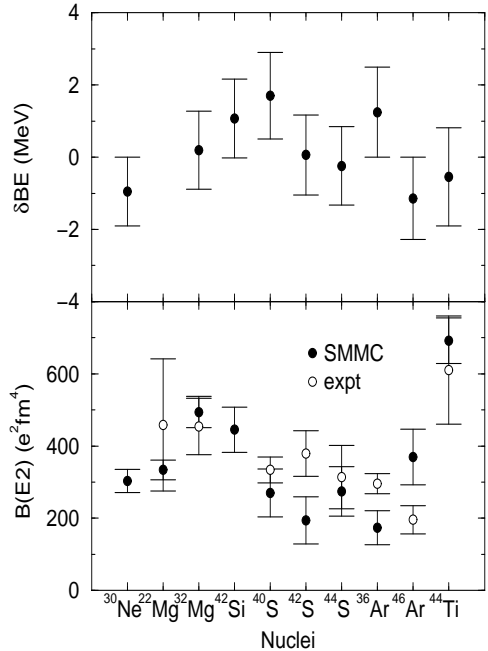,height=3.5in,width=2.5in}}
\caption{Top: Difference between theoretical and experimental
binding energies for the $sd$-$pf$-shell nuclei studied in this work.
Bottom: Experimental and theoretical $B(E2)$ values.
\label{fig_cm}}
\end{figure}

Most of the measured observables can be calculated within the SMMC
framework.  It is well known that in {\it deformed} nuclei the total
$B(E2)$ strength is almost saturated by the $0^+_{gs} \rightarrow
2_1^+$ transition (typically 80\% to 90\% of the strength lies in this
transition).  Thus the total strength calculated by SMMC should only
slightly overestimate the strength of the measured transition.  In
Fig.~\ref{fig_cm}
the SMMC computed $B(E2, total)$ values are
compared to the experimental $B(E2; 0^+_{gs} \rightarrow 2^+_1)$
values.  Reasonable agreement with experimental data across the
space is obtained when one chooses effective charges of $e_p=1.5$ and
$e_n=0.5$. Using these same effective charges,
the USD values for the $B(E2,0_{gs}^+ \rightarrow 2_1^+)$
of the $sd$-shell nuclei $^{32}$Mg and $^{30}$Ne are
177.1 and 143.2 $e^2$fm$^4$, respectively, far lower than the full
$sd$-$pf$ calculated and experimental values.
All of the theoretical
calculations require excitations to the $pf$-shell before reasonable
values can be obtained.  We note a general agreement among all
calculations of the $B(E2)$ for $^{46}$Ar, although they are typically
larger than experimental data would suggest. We also note a somewhat
lower value of the $B(E2)$ in this calculation as compared to
experiment and other theoretical calculations in the case of $^{42}$S
\cite{r:retamosa}.

Also shown in Fig.~\ref{fig_cm} are the differences between experimental
and theoretical binding energies for nuclei in this region. Agreement
is quite good overall. Further details of the interaction and results
may be found in \cite{dean_cm}.

\subsection{Electron capture rates for Fe-region nuclei}

\begin{table}
\caption
{Comparisons of the SMMC electron capture rates  with the total
($\lambda_{\rm ec}$) and
partial Gamow-Teller ($\lambda^{\rm GT}_{\rm ec}$) rates as given in
Ref.~\protect\cite{Aufderheide}. Physical conditions at
which the comparisons were made are
$\rho_7=5.86$, $T_9=3.40$, and $Y_e=0.47$
for the upper part of the table, and
$\rho_7=10.7$, $T_9=3.65$, and $Y_e=0.455$ for the lower part.
}
\begin{center}
\begin{tabular}{|c|c|c|c|}
\hline
Nucleus
& $\lambda_{\rm ec}$ (sec$^{-1}$)
& $\lambda_{\rm ec}$ (sec$^{-1}$)
& $\lambda_{\rm ec}^{GT}$ (sec$^{-1}$) \\
&   (SMMC)  &
(Ref.~\protect\cite{Aufderheide}) &
(Ref.~\protect\cite{Aufderheide})  \\
\hline
$^{55}$Co &  3.89E-04  &   1.41E-01 & 1.23E-01 \\
$^{57}$Co &  3.34E-06  &   3.50E-03 & 1.31E-04 \\
$^{55}$Fe &  1.20E-08  &   1.61E-03 & 1.16E-07 \\
$^{56}$Ni &  3.47E-02  &   1.60E-02 & 6.34E-03 \\
$^{58}$Ni &  1.01E-03  &   6.36E-04 & 4.04E-06 \\
$^{60}$Ni &  7.39E-05  &   1.49E-06 & 4.86E-07 \\
\hline
$^{59}$Co &  3.44E-07  &   2.09E-04 & 6.37E-05 \\
$^{57}$Co &  2.06E-05  &   7.65E-03 & 3.69E-04 \\
$^{55}$Fe &  1.07E-07  &   3.80E-03 & 5.51E-07 \\
$^{56}$Fe &  9.80E-06  &   4.68E-07 & 6.60E-10 \\
$^{54}$Fe &  3.84E-04  &   9.50E-04 & 3.85E-06 \\
$^{51}$V  &  1.06E-06  &   1.24E-05 & 9.46E-09 \\
$^{52}$Cr &  1.32E-04  &   2.01E-07 & 1.59E-10 \\
$^{60}$Ni &  3.61E-04  &   7.64E-06 & 2.12E-06 \\
\hline
\end{tabular}
\end{center}
\end{table}

The impact of nuclear structure on
astrophysics has become increasingly
important, particularly
in the fascinating, and presently unsolved, problem of type-II supernovae
explosions.
One key ingredient of the precollapse scenario is the electron-capture
cross section on nuclei \cite{Bethe90,Aufderheide}. An important
contribution to electron-capture cross sections
in supernovae environments is the Gamow-Teller (GT) strength distribution.
This strength distribution, calculated in SMMC using Eqs.~(3 and 4) above,
is used to find the energy-dependent cross section for electron
capture. In order to obtain the electron-capture rates,
the cross section is then folded with
the flux of a degenerate relativistic electron gas \cite{Dean98}.
Note that the Gamow-Teller distribution is calculated at
the finite nuclear temperature which, in
principle, is the same as the one for the electron gas.

It is important to calculate the GT strength distributions
reasonably accurately for both the total strength and the position
of the main GT peak in order to have a quantitative estimate for the
electron-capture rates. For astrophysical purposes, calculating the
rates to within a factor of two is required.
We concentrate here on mid-$fp$-shell results for the electron-capture
cross sections\cite{Dean98}.
The Kuo-Brown interaction \cite{KB}, modified in the monopole
terms by Zuker and Poves \cite{KB3}, was used throughout these $pf$-shell
calculations. This interaction reproduces quite nicely the ground- and
excited-state properties of mid-$fp$-shell nuclei \cite{Caurier1,Langanke95},
including the total Gamow-Teller strengths and distributions, where
the overall agreement between theory and experiment \cite{radha97} is
quite reasonable.
The SMMC technique allows one to probe the
complete $0\hbar\omega$ $fp$-shell region
without any parameter adjustments
to the Hamiltonian, although the Gamow-Teller
operator has been renormalized by the standard factor of 0.8.

Do the electron-capture rates presented here
indicate potential implications for
the precollapse  evolution of a type II supernova? To make a
judgment on this
important question, one should compare
in Table I the SMMC rates for selected
nuclei with those currently used in collapse calculations
\cite{Aufderheide}. For the comparison, we choose the same physical
conditions as assumed in Tables 4--6 in Aufderheide
{\it et al. }\cite{Aufderheide}.
Table I also lists the partial electron-capture rate which has been
attributed to Gamow-Teller transitions \cite{Aufderheide}.
Note that for even-parent nuclei, the present rate
approximately agrees with the currently recommended {\it total} rate. A
closer inspection, however, shows significant differences between the
present rate and the one attributed to the Gamow-Teller transition in
Aufderheide {\it et al.}~\cite{Aufderheide}.
The origin of this discrepancy
is due to the fact that Fuller {\it et al.} \cite{FFN}
places the Gamow-Teller
resonance for even-even nuclei systematically at too high an excitation
energy. This shortcoming has been corrected
in Fuller {\it et al.} \cite{FFN} and Aufderheide {\it et al.}
\cite{Aufderheide} by adding an experimentally known
low-lying strength in addition to
the one attributed to Gamow-Teller transitions.
However, the overall good agreement between the SMMC results for
even-even nuclei and the recommended rates indicates that the SMMC
approach also accounts correctly for this low-lying strength. This has already
been deduced from the good agreement between SMMC Gamow-Teller
distributions and data including the low-energy regime \cite{radha97}.

Thus, for even-even nuclei, the SMMC approach is able to
predict the {\it total} electron-capture rate rather reliably, even if
no experimental data are available. Note that the SMMC rate is
somewhat larger than the recommended rate for $^{56}$Fe and $^{60}$Ni.
In both cases, the experimental Gamow-Teller distribution is known and
agrees well with the SMMC results \cite{radha97}. While the proposed
increase of the rate for $^{60}$Ni is not expected to have noticeable
influence on the pre-collapse evolution, the increased rate for $^{56}$Fe
makes this nucleus an important contributor in the change of $Y_e$
during the collapse (see Table 15 of Aufderheide {\it et al.}
\cite{Aufderheide}).

For electron capture on odd-$A$ nuclei, observe that
the SMMC rates, derived from the Gamow-Teller
distributions, are significantly smaller than the recommended total rate.
This is due to the fact that for
odd-$A$ nuclei the Gamow-Teller transition peaks at rather high
excitation energies in the daughter nucleus. The electron-capture rate
on odd-$A$ nuclei is therefore carried by weak transitions at low
excitation energies. Comparing the SMMC rates to those attributed to
Gamow-Teller transitions in Fuller {\it et al.}\cite{FFN} and
Aufderheide {\it et al.}\cite{Aufderheide} reveals that
the latter have been, in general, significantly overestimated, which is
caused by the fact that the position of the Gamow-Teller
resonance is usually put at too low excitation energies in the daughter.
The SMMC calculation implies that the Gamow-Teller transitions should not
contribute noticeably to the electron-capture rates on odd-$A$ nuclei at
the low temperatures studied in Tables 14--16 in
Aufderheide {\it et al.}\cite{Aufderheide}.
Thus, the rates for odd-$A$ nuclei given in these tables should generally
be replaced by the non-Gamow-Teller fraction.

\subsection{Pair correlations in nuclei}

We would now like to turn to the subject of pair correlations in nuclei,
and calculations aimed at their understanding.
Nuclei near N$=$Z offer a unique place to study proton-neutron pairing,
particularly in the isospin T=1 channel. In fact, most heavy odd-odd,
N$=$Z nuclei beyond $^{40}$Ca have total spin J$=$0, T$=$1 ground states.
Theoretical studies have shown that many of these nuclei have enhanced
T$=$1 proton-neutron correlations when compared to their even-even
counterparts. These correlations are, to a lesser extent,
present in even-even systems, but tend to decrease as one moves away
from N$=$Z. In at least one nucleus in the mass 70 region, $^{74}$Rb,
there is experimental evidence for a ground state $T=1$ band \cite{rudolph}.

\begin{figure}
{\hskip 0.5in \psfig{file=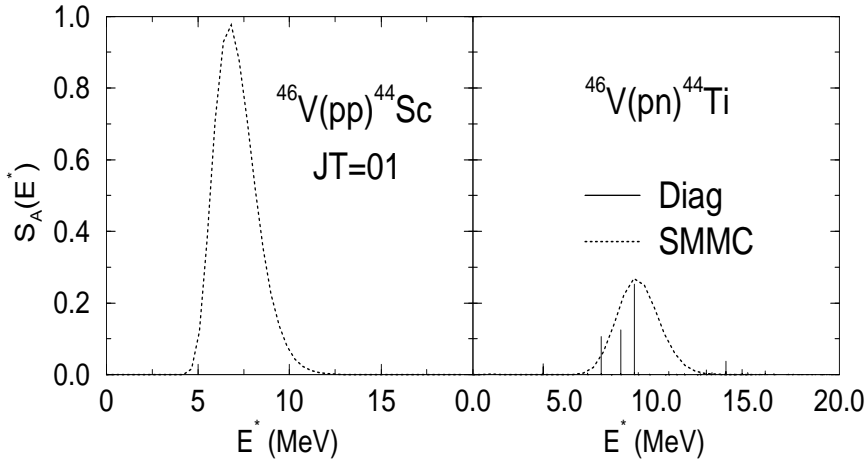,height=2.5in,width=4.5in}}
\caption{Left: proton pair
strength distribution for $^{46}$V. Right:
proton-neutron (T=1) pair strength distribution. SMMC: thick line;
direct diagonalization: impulses.
\label{foobar:fig1}}
\end{figure}

Experimentally, pair correlations can best be
measured by pair transfer on nuclei. Although total
cross sections are typically underpredicted when one
employs spectroscopic factors computed from the shell model,
relative two-nucleon spectroscopic factors within one
nucleus are more reliable. Therefore, it is necessary
for one to calculate
and measure pair transfer from both the ground and
excited states in a nucleus.

The SMMC method may be used to calculate
the strength distribution of the pair
annihilation operator $A_{JTT_z}$, as defined in
Koonin {\it et al.}\cite{Koonin97}.
The total strength of these pairing operators, i.e. the
expectation  $\langle A^\dagger_{JT} A_{JT}\rangle$, has been
studied previously
as a function of mass, temperature \cite{langanke96,langanke97}, and
rotation \cite{dean97}. We would like to briefly present here
the strength distributions of
the pair operators as calculated in SMMC.
The strength distribution for the pair transfer spectroscopic
factors is proportional to $\langle A-2 \mid A_{JT}\mid A\rangle$
and is calculated by the inversion of Eq.~(4).

In future work, we will discuss the strength distributions in detail.
Here we would like to briefly conclude
by demonstrating that the SMMC results and the
direct diagonalization results agree very nicely for the proton
pair strength distributions in the ground state of $^{46}$V.
This is demonstrated in the left panel of
Fig. 2. Shown in the right panel is the isovector proton-neutron
pairing strength distribution with respect to the
daughter nucleus. Notice that the overall strength
is much larger in the proton-neutron channel, as discussed previously in
Langanke {\it et al.} \cite{langanke97}, and that the peak
is several MeV lower in excitation relative to the like-particle
channel. In both cases the strength distribution in $^{46}$V differs
significantly from that found in $^{48}$Cr, where
one finds that the dominant component is a ground-state to ground-state
transition involving mainly particles in the $0f_{7/2}$ single-particle
state. In both odd-odd N$=$Z channels, the distribution is fairly highly
fragmented.

\subsection{Rare earth nuclei}

We have recently applied SMMC techniques to survey rare-earth nuclei
in the Dy region. This extensive study formed the thesis topic of
J.A. White \cite{jaw}, whose goal was to
examine how the phenomenologically motivated ``pairing-plus-quadrupole''
interaction compares in exact shell-model solutions with other methods.
We also examined how the shell-model solutions compare with
experimental data;  static path approximation (SPA)
calculations were also performed.  There have been efforts
recently by others to use SPA calculations, since they are simpler and
faster (see \cite{Rossignoli,Rossignoli2} as examples).
However, we found that SPA results are not consistently good.
This study was also designed to investigate whether the
phenomenological pairing-plus-quadrupole-type interactions can be used
in exact solutions for large model spaces, and whether the interaction
parameters require significant renormalization when using SPA.

We discuss here one particular aspect of that work,
namely level density calculations. Details may be found in
\cite{JW98}.
We used the Kumar-Baranger Hamiltonian with parameters appropriate
for this region. Our single-particle space included the 50-82 subshell
for the protons and the 82-126 shell for the neutrons.
While several interesting aspects of these systems were studied in
SMMC, we limit our discussion here to the level densities obtained for
$^{162}$Dy.

SMMC is an excellent way to calculate level densities.  $E(\beta)$ is
calculated for many values of $\beta$ which determine the partition
function, $Z$, as
\begin{equation}
\ln[Z(\beta)/Z(0)]=-\int_0^\beta d\beta'E(\beta')
\end{equation}
$Z(0)$ is the total number of available states in the space.  The
level density is then computed as an inverse
Laplace transform of $Z$.  Here, the last step is performed with a
saddle point approximation with $\beta^{-2}C\equiv -dE/d\beta$:
\begin{eqnarray}
S(E)& = & \beta E + \ln Z(\beta) \\
\rho(E)&=&(2\pi\beta^{-2}C)^{-1/2}\exp(S)
\label{eq:rho}
\end{eqnarray}
SMMC has been used recently to calculate level density in
iron region nuclei~\cite{density}, and here we demonstrate its
use in the rare-earth region.

\begin{figure}
{\hskip 0.5in \psfig{file=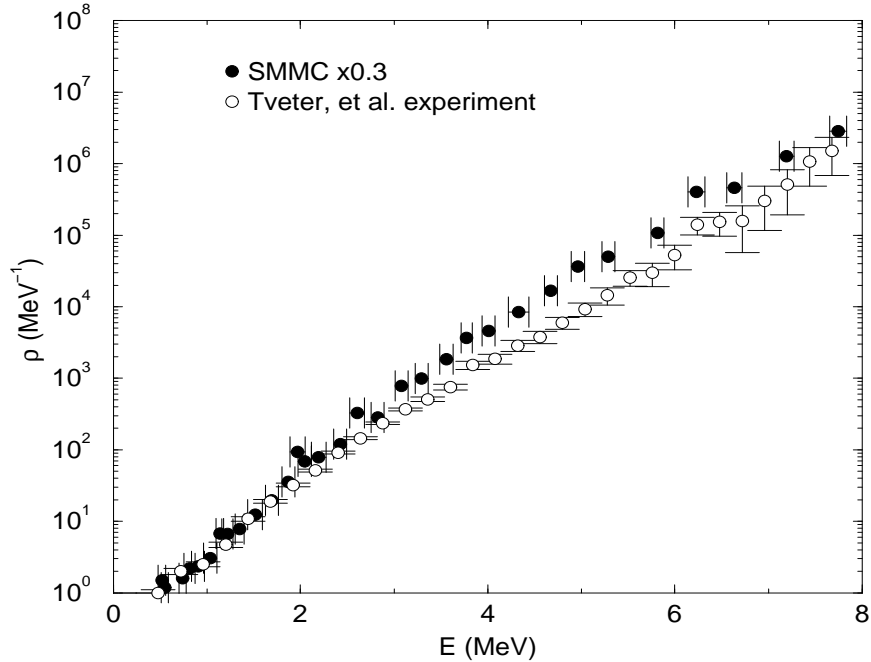,height=3.5in,width=4.5in}}
\caption{SMMC density vs. experimental
data in $^{162}$Dy.
\label{fig:dy162}}
\end{figure}

The comparison of SMMC density in $^{162}$Dy with the Tveter et
al.~\cite{Tveter96} data is displayed in Fig.~\ref{fig:dy162}.
The experimental
method can reveal fine structure, but does not determine the absolute
density magnitude.  The SMMC calculation is scaled by a factor to
facilitate comparison.  In this case, the factor has been chosen to
make the curves agree at lower excitation energies.  From 1-3 MeV,
the agreement is very good.  From 3-5 MeV, the SMMC density
increases more rapidly than the
data.  This deviation from the data cannot be accounted for by
statistical errors in either the calculation or measurement.  Near 6
MeV, the measured density briefly flattens before increasing and this
also appears in the calculation, but the measurement errors are larger
at that point.

The measured density includes all states
included in the theoretical calculation plus some others, so that one
would expect the measured density to be greater than or equal to the
calculated density and never smaller.  We may have instead chosen our
constant to match the densities for moderate excitations and let the
measured density be higher than the SMMC density for lower energies
(1-3 MeV). Comparing structure between SMMC and data is difficult
for the lowest energies due to statistical errors in the calculation
and comparison at the upper range of the SMMC calculation, i.e.,
$E\approx15$ MeV, is unfortunately impossible since the data only
extend to about $8$ MeV excitation energy.

\section{Conclusions}

In these Proceedings, we have used four specific examples (there are several
others) for which the SMMC calculations have proven very useful in
understanding the properties of nuclei in systems where the number
of valence particles prohibits the use of more traditional approaches.
The method has proven to be a valuable tool for furthering our understanding
of nuclear structure and astrophysics. Continued developments
in both creating useful interactions and shell-model technology
should continue to enhance our ability to understand nuclei far from
stability in the coming years.

\section*{Acknowledgments}
Oak Ridge National Laboratory (ORNL)
is managed by Lockheed Martin Energy Research Corp. for the
U.S. Department of Energy under contract number DE-AC05-96OR22464.
This work was supported in part through grant DE-FG02-96ER40963 from the
U.S. Department of Energy. This work was supported in part by
the National Science Foundation,
Grants No. PHY-9722428, PHY-9420470, and PHY-9412818.

\end{document}